\documentclass[reprint,amsmath,amssymb,aps,prb,groupedaddress]{revtex4-1}

\usepackage{graphicx}
\usepackage{dcolumn}
\usepackage{bm}

\begin{document}

\title{Possible structural and bond reconstruction in 2D ferromagnetic semiconductor VSe$_{2}$ under uniaxial stress}

\author{Bo-Wen Yu}
\affiliation{Beijing National Laboratory for Condensed Matter Physics, Institute of Physics, Chinese Academy of Sciences, Beijing 100190, China}
\affiliation{School of Physical Sciences, University of Chinese Academy of Sciences, Beijing 100190, China}
\author{Bang-Gui Liu}\email{bgliu@iphy.ac.cn}
\affiliation{Beijing National Laboratory for Condensed Matter Physics, Institute of Physics, Chinese Academy of Sciences, Beijing 100190, China}
\affiliation{School of Physical Sciences, University of Chinese Academy of Sciences, Beijing 100190, China}

\date{\today}

\begin{abstract}
Two-dimensional (2D) semiconducting transition metal dichalcogenides have been used to make high-performance electronic, spintronic, and optoelectronic devices. Recently, room-temperature ferromagnetism and semiconducting property were found in 2D VSe$_2$ nanoflakes (mechanically exfoliated onto silicon substrates capped with a oxide layer) and are attributed to the stable 2H-phase of VSe$_2$ in the 2D limit. Here, our first-principles investigation show that a metastable semiconducting H' phase can be formed from the H VSe2 monolayer and some other similar when these 2D H-phase materials are under uniaxial stress or uniaxial strain. For the uniaxial stress (uniaxial strain) scheme, the H' phase will become lower in total energy than the H phase at the transition point. The calculated phonon spectra indicate the dynamical stability of the H' structures of VSe$_2$, VS$_2$, and CrS$_2$, and the path of phase switching between the H and H' VSe$_2$ phases is calculated. For VSe$_2$, the H' phase has stronger ferromagnetism and its Currier temperature can be substantially enhanced by applying uniaxial stress or strain.
Spin-resolved electronic structures, energy band edges, and effective carrier masses for both of the H and H' phases can be substantially changed by the applied uniaxial stress or strain, leading to huge effective masses near the band edge of the strained H' phase.
Analysis indicated that the largest bond length difference between the H' and H phases can reach -19\% for the Se3-Se3' bond, and there is noticeable covalence for the Se3-Se3' bond, which switches the valence of the nearby V atoms, leading to the enhanced ferromagnetism. Therefore, structural and bond reconstruction can be realized by applying uniaxial stress in 2D ferromagnetic H VSe$_2$ and some other similar. These can be useful to seeking more insights and phenomena in such 2D materials for potential applications.
\end{abstract}


\maketitle

\section{Introduction}

Since the advent of graphene, two-dimensional (2D) materials have attracted huge attention around the world and great effort has been made to achieve high-performance electronic, spintronic, and optoelectronic devices\cite{r1,r2,r3,r4,r5,r6,r7}. Transition metal dichalcogenides (TMDs) not only assume three-dimensioanl (3D) layered crystal structures, but also form  2D monolayer materials, with various physical properties different from the corresponding 3D bulk materials\cite{r4,r5,r6,r7}. There are two kinds of common structures for 2D TMDs: T structure and H structure.
Among them, MoX$_2$ and WX$_2$ (X=S, Se, Te) are the most famous, and they have been explored intensively and various interesting devise have been realized from them\cite{r4,r5,r6,r7,r77,r7a,r7b}. In addition, transition metal can be replaced by rare earth atoms and chalcogen by halogens in these 2D structures, producing various interesting 2D materials\cite{re1,re2,re3,re4}.
On the other hand, 2D VX$_{2}$ (X=S, Se, Te) become very attractive. 2D VS$_2$\cite{r8,r9,r10}, 2D VSe$_2$\cite{r11,r12,r13,r14}, and VTe$_2$\cite{r15,r16,r17,r18} were widely studied to seek ideal material platforms for the materials of electrodes, possible room-temperature 2D ferromagnetism, etc. It was shown that a structural phase transition of multilayer VSe2 from 1T metallic phase to 2H semiconductive phase can be accomplished through annealing at 650 K and the 2H-phase is more thermodynamically favorable than the 1T-phase at the 2D limit\cite{adTr}. Recently, 2H VSe$_2$ single crystals were grown and VSe$_2$ nanoflakes were mechanically exfoliated onto silicon substrates capped with a oxide layer, and then room-temperature ferromagnetic semiconductor was found in the nanoflakes and attributed to the 2H-phase of VSe$_2$ in the 2D limit\cite{r13}. Further exploration of H VSe$_2$ monolayer is highly desirable to obtain more insights and phenomena in order to seek potential applications for electronics, spintronics, and valleytronics\cite{r7,r7b}.

Here, we study on the effects of uniaxial strain or stress on 2D H VSe$_2$ and other monolayers by the first-principles calculations and find a metastable H' structure of them when uniaxial strain or stress is applied on the H phase along the armchair (y) direction. This reconstructed H' phase will be lower in total energy than the H phase when the uniaxial strain or stress reaches a transition point. We investigate its structural stability, phase transition barrier, ferromagnetism, electronic structures, chemical bonds, etc. It is a ferromagnetic semiconductor, and its ferromagnetism and Currier temperature can be substantially enhanced by applying uniaxial strain or stress. Our analyses show that the H' phase can be seen to form from the H phase through strain-driven structural and bond reconstruction.  Computational methods and more detailed results will be presented in the following.

\section{Methodology}

All first-principles calculations are performed with the projector-augmented wave (PAW) method within the density functional theory\cite{r19}, as implemented in the Vienna Ab-initio simulation package software (VASP) \cite{r20}. The generalized gradient approximation (GGA) by Perdew, Burke, and Ernzerhof (PBE)\cite{r21} is used as the exchange-correlation functional. The  self-consistent calculations are carried out with a $\Gamma$-centered ($20\times 5\times 1$) Monkhorst-Pack grid\cite{r22}. The kinetic energy cutoff of the plane wave is set to 450 eV. The convergence criteria of the total energy and force are set to 10$^{-7}$ eV and 0.01 eV/\AA{}. The spin-orbit coupling (SOC) is used in the calculation of band structure and magnetocrystalline anisotrpy energy. The nudged elastic band method (NEB) \cite{r23} is used to find the minimum energy path of the phase transitions.  The  Hubbard-U term \cite{r25} ($U=2$eV) is considered in the DFT+U scheme to improve energy band description. The Phonopy package \cite{r26} is used to calculate the phonon spectra.

\begin{figure}
    \includegraphics[width=\columnwidth]{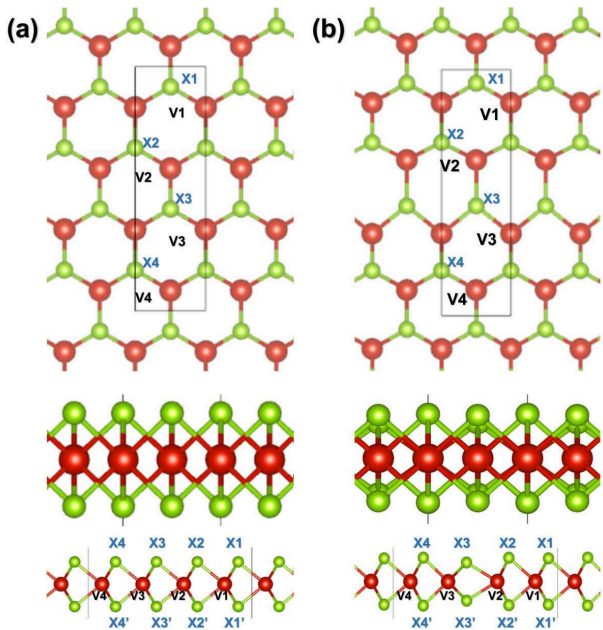}
    \caption{\label{fig1} The H structure (a) and the H' structure (b) of VSe$_2$. The top row shows the top view (the xy plane), the middle row describes the side view (the yz plane) along the x axis, and the bottom row demonstrates the side view (the xz plane) along the y axis. V$_i$ denotes different V atoms, and X$_i$ (X$^\prime_i$) describes different Se atoms in the upper (lower) Se plane.}
\end{figure}

\section{Result and discussion}

\subsection{Effects of uniaxial strain and stress}

The structure of H-phase VX$_2$ (X = Se, S, Te) is demonstrated in Fig. \ref{fig1}. The high symmetry of the H structure will be destroyed if uniaxial tensile stress is applied in the x direction (zigzag direction) or y direction (armchair direction). According to the Poisson effect, if there is a strain in the y axis, for example $\eta_y$, there will be a perpendicular strain $\eta_x$ determined by total energy optimization (or the zero stress condition $\sigma_x=0$). This realizes the y-axis uniaxial stress $\sigma_y=\partial E /A\partial\eta_y$ (with $\sigma_x=0$), where $E$ is the total energy, $\eta_y$ is the y-axis strain and $A$ is the unit-cell area in the xz plane of our computational model\cite{zhangsh}. On the other hand, for convenience, we can assume an uniaxial strain in the y direction, $\eta_y$, which implies $\eta_x=0$. Without the uniaxial strain or stress, the H structure has C$_3$-centered symmetry (P$\bar{3}$m1 space group for our computational model).  Under the uniaxial strain $\eta_y$ or the uniaxial stress $\sigma_y$, the C$_3$ symmetry will be replaced by a mirror symmetry and thus we choose the rectangle cell marked in Fig. \ref{fig1}(a). Therefore, we can manipulate the H-phase VX$_2$ by using uniaxial strain or uniaxial stress.

\begin{table}[h]
    \caption{\label{table1}%
        The energy difference $\Delta E$ between the H' and  H structures of VX$_2$ (X = Se, Te, and S) . $E_H$ and $E_{H'}$ are the total energies of the H and H' structures, respectively. }
    \begin{ruledtabular}
        \begin{tabular}{cccc}
            \textrm{System}               &
            \textrm{$E_H$ (eV/f.u.)}    &
            \textrm{$E_{H'}$ (eV/f.u.)} &
            \textrm{$\Delta E$ (eV/f.u.) }                              \\
            \colrule
            VSe$_2$                     & -15.8347 & -15.9656 & 0.1309 \\
            VTe$_2$                     & -17.5369 & -17.6866 & 0.1497 \\
            VS$_2$                      & -14.2328 & -14.2434 & 0.0106 \\
        \end{tabular}
    \end{ruledtabular}
\end{table}

\begin{table}[h]
    \caption{\label{table2}%
        The equilibrium lattice constants ($a$ and $b$), magnetic moment ($M$), and V-X bond lenthes ($d_\text{V-X}$) of the H' VX$_2$ (X = Se, Te, and S) (with the local lowest energy)}
    \begin{ruledtabular}
        \begin{tabular}{cccccc}
            Name  & $a$ (\AA) & $b$ (\AA) & $M$ ($\mu_B$) & $d_\text{V-X}$ (\AA) & $d_\text{X-X'}$ (\AA)\\ \colrule
            VSe$_2$   & 3.42   & 12.14    & 6.000 & 2.47-2.62 & 2.66-3.29 \\
            VTe$_2$   & 3.61   & 13.19    & 6.000 & 2.68-2.83 & 3.01-3.67 \\
            VS$_2$    & 3.25   & 11.58    & 6.000 & 2.37-2.46 & 2.46-3.05 \\
        \end{tabular}
    \end{ruledtabular}
\end{table}

\begin{table}[h]
    \caption{\label{table3}%
        The bond lengthes $l_b$ (\AA) and angles $\alpha_b$ ($^\circ$) of the H and H' phases of VSe$_2$ at $b^\prime_0$ ($\eta^\prime_y=0$,$\eta_y$=4.3\%). The V-Se and V-Se' share the same bond length.}
    \begin{ruledtabular}
        \begin{tabular}{ccc|ccc}
        $l_b$ & \textrm{H' phase} & \textrm{H phase} & $\alpha_b$ & \textrm{H' phase} & \textrm{H phase}          \\
            \colrule
            V1-Se1  & 2.53 & 2.53 & Se1-V1-Se1' & 78.31  & 77.85\\
            V1-Se2  & 2.55 & 2.56& Se2-V1-Se2' & 79.53   & 76.52\\
            V2-Se2  & 2.52 & 2.53 & Se2-V2-Se2' & 80.88   & 77.85\\
            V2-Se3  & 2.62 & 2.56& Se3-V2-Se3' & 60.95   & 76.52\\
            V3-Se3  & 2.55 & 2.53 & Se3-V3-Se3' & 62.71   & 77.85 \\
            V3-Se4  & 2.47 & 2.56& Se4-V3-Se4' & 83.72   & 76.52\\
            V4-Se4  & 2.56 & 2.53 & Se4-V4-Se4' & 80.21   & 77.85\\
            V4-Se1  & 2.52 & 2.56& Se1-V4-Se1' & 78.74   & 76.52\\ \hline
            Se1-Se1' &3.19  & 3.18& & &   \\
            Se2-Se2' &3.26  & 3.18& & &   \\
            Se3-Se3' &2.65  & 3.18& & &   \\
            Se4-Se4' &3.19  & 3.18& & &   \\
        \end{tabular}
    \end{ruledtabular}
\end{table}

\begin{figure}
    \includegraphics[width=0.8\columnwidth]{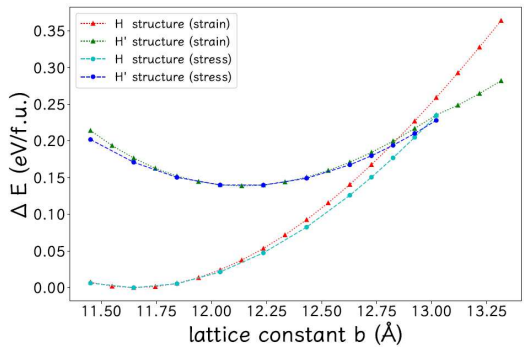}
    \caption{\label{fig2} The effects of the uniaxial strain and uniaxial stress on the total energy values of the H'  and H  phases of the VSe$_2$. The total energy difference between the equilibrium H' and H phases is 0.13 eV per formula unit, and the corresponding equilibrium lattice constants are $b_0^\prime$=12.14 \AA{} and  $b_0$=11.65 \AA{}, respectively.}
\end{figure}

We shall use strain $\eta_y$ or lattice constant $b$ to characterize the VX$_2$ in the cases of uniaxial strain and uniaxial stress in the y axis. The uniaxial y strain implies $\eta_x=0$ (in the x axis), but the uniaxial y stress means $\sigma_x=0$ (in the x axis).
We show in Fig. \ref{fig2} the total energies as functions of $b$ for both uniaxial strain and uniaxial stress. It is clear that the H-phase corresponds to the globally lowest total energy and for a given $b$ value the total energy under the uniaxial stress is lower than that under the uniaxial strain.
More interestingly, as shown in Figs. \ref{fig1}(b) and \ref{fig2}, our study shows that there is a metastable structure H' when we apply the uniaxial strain to the H- structure of VX$_2$ (X = Se, S, and Te). For the semiconductive H'-VSe$_2$ the total energy is higher by 0.13 eV than the corresponding H phase, which is similar to a predicted metastable metallic 1T-type phase\cite{ad1T}. The predicted lattice constants and total energies for all the VX$_2$ are summarized in Table \ref{table1}. Most noticeably, the top view of the bond lengthes are not uniformly stretched along the y direction, as shown in Fig. \ref{fig1}(b) and Table \ref{table2}. Actually, this is caused by the shortened X$_3$-X$^\prime_3$ bond length with respect to that of other X pairs, as shown in Table \ref{table3}. Hence, the H' structure can be seen as a reconstructed phase of the H phase.

For VSe$_2$, the H' structure has the lowest total energy when the lattice constant  (y-direction) is $b_0^\prime$=12.14 \AA{}, smaller by 4.3\% than the equilibrium lattice constant $b_0=11.64$ \AA{} of the H-phase structure. As shown in Fig. \ref{fig2}, the H' phase will be favored in total energy when the $b$ reaches $b_{c1}$=12.82 \AA{} (the uniaxial strain $\eta_{yc1}$=10.1\% with respect to $b_0$). In the uniaxial stress case, when the lattice constant $b$ reaches $b_{c2}$=13.12\AA{} ($\eta_{yc2}$=12.7\%), the H structure will be higher in total energy than the H' structure. It should be pointed out that the transition points $b_{c1}$=12.82 \AA{} for the uniaxial strain case and $b_{c2}$=13.12\AA{} for the uniaxial stress case correspond to $\eta^\prime_{yc1}$=5.6\% and $\eta^\prime_{yc2}$=8.1\% with respect to the H' equilibrium lattice constant $b_0^\prime$, respectively. These imply that when $b$ becomes larger than $b_{c1}$ for the uniaxial strain case (or $b_{c2}$ for the uniaxial stress case), the VSe$_2$ can transit from the H phase to the H' phase, which can be seen to realize a structural and bond reconstruction in the monolayer VSe$_2$.

\subsection{Structural stability and phase switching}

\begin{figure}[h]
    \centering
    \includegraphics[width=0.48\textwidth]{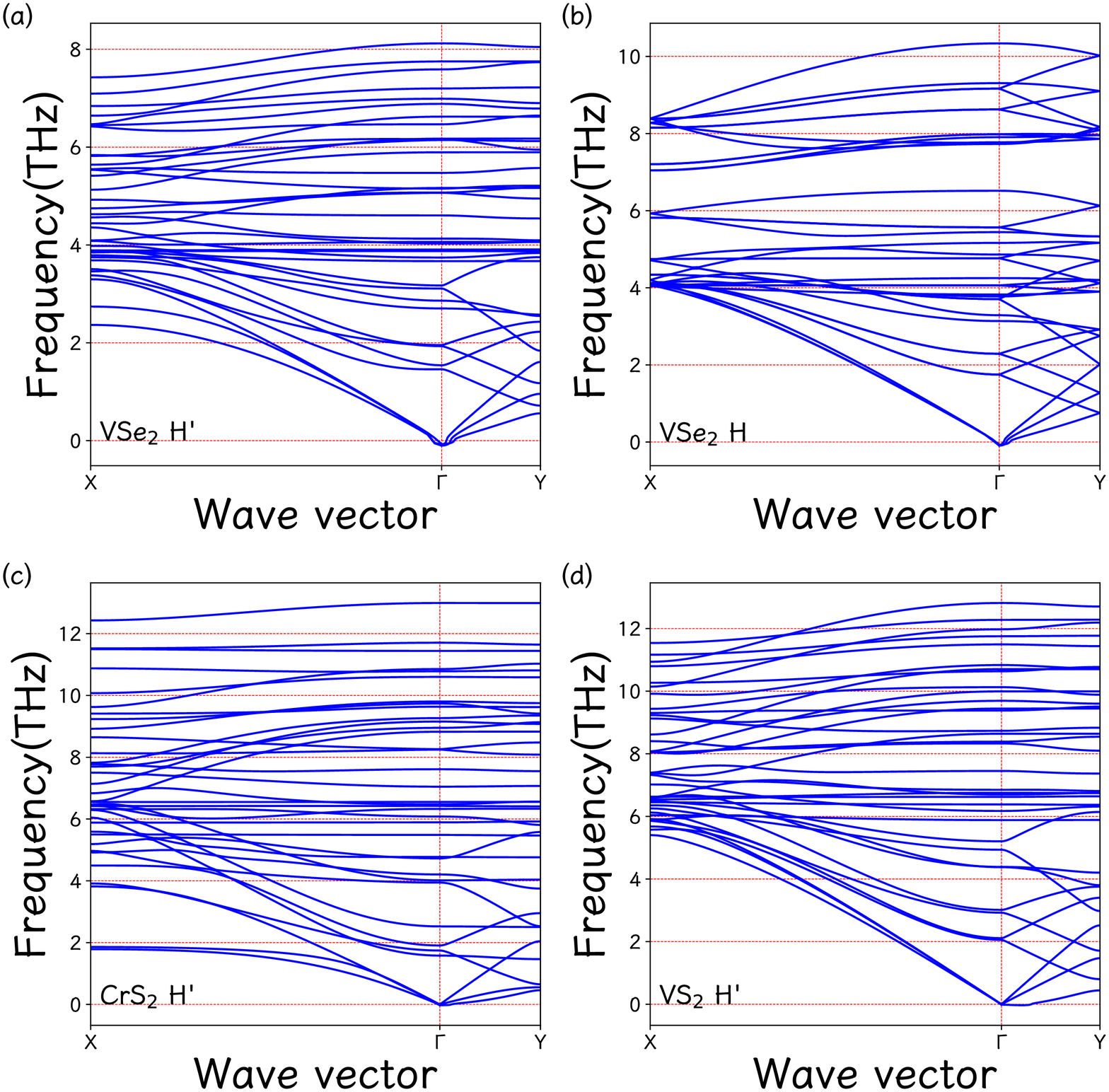}
    \caption{\label{fig3} The phonon spectra of the H'-phase VSe$_2$ (a), H-phase VSe$_2$ (b), H'-phase CrS$_2$ (c), and H'-phase VS$_2$.}
\end{figure}

More exploration has been conducted to seek more similar 2D structures. It is found that both CrS$_2$ and CrSe$_2$ also have the H' structure under similar uniaxial strain. In order to ensure the existence of the new structures, it is necessary to conduct the phonon dispersion calculations for the 2D structures. In Fig. \ref{fig3} we compare the phonon spectra of the H' and H structures of VSe$_2$ and present phonon spectra of H'-phases of CrS$_2$ and VS$_2$. These results show that the H'-phase is dynamically stable  for  VSe$_2$, VS$_2$, and CrS$_2$. The result of VTe$_2$ can also show stability within the calculation accuracy. Our calculation with CrSe$_2$ indicates that its total energy is lower than that of the corresponding H structure, while it is challenging to obtain good phonon spectra, which means that there might be a more stable structure for CrSe$_2$ under uniaxial strain.

\begin{figure}[h]
    \includegraphics[width=0.8\columnwidth]{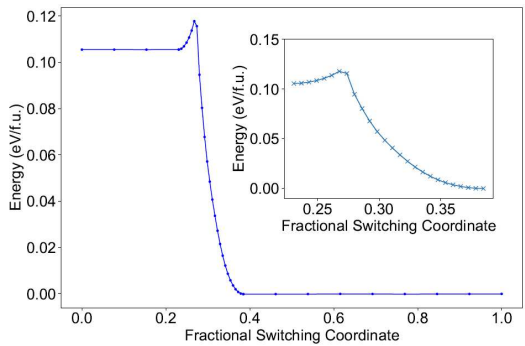}
    \caption{\label{fig4} A possible minimum energy path of the phase switching between the H structure and the H' structure of VSe$_2$. The energy barrier is 0.012 eV.}
\end{figure}

As for the total energy difference between the H structure and H' structure of the VX$_2$ (X = Se, Te, and S) shown in Table~\ref{table1}, it is 0.1309 eV per formula unit in the case of VSe$_2$. As a rule, the H structure has the lower total energy than the H' structure, which indicates that the H' phase is metastable and could be realized by applying uniaxial strain or stress.
It is also essential to calculate the energy barrier of the phase switching between the H and H' structures. One of the minimum energy path can be calculated by the nudged elastic band (NEB) method\cite{r23}, as is shown in Fig.~\ref{fig4}. The energy barrier is 0.012 eV, which indicates that the H' structure of VSe$_2$, when achieved, will be stable at low temperature.

\subsection{Ferromagnetism and Currier temperature}

Because there is a single d electron per V atom left after bonding with the $4+$ valence, a magnetic moment of 1.000 $\mu_B$ per V atom will be formed in terms of first-principle calculations\cite{r27,r28,r29}.  According to the symmetry, the H structure without any uniaxial strain or stress has C$_3$ symmetry. For H-phase VX$_2$, the magnetic moment is $4.000\mu_B$ per unit cell, which reflects that there are 4 V atoms in the unit cell and a V atom assumes chemical valence $4+$. For H'-phase VX$_2$, the magnetic moment is enhanced to $6.000\mu_B$, which can be explained by noting that the shortened X$_2$-X$^\prime_2$ bond length makes the neighboring V atoms assume a different valence $3+$ so that the two V atoms per unit cell contribute 2.000$\mu_B$ more than in the H phase. The magnetocrystalline anisotrpy energy (MAE) calculated by comparing the total energies along different directions is very important to the magnetization behavior. For VSe$_2$, the total energy values of the H and H' structures under the zero strains and the two transition strains ($\eta_{yc1}$ and $\eta_{yc2}$), with the magnetic moment orienting in the [100], [010], and [001] directions,  are summarized in Table~\ref{table4}. It is clear that the x (zigzag) axis is the easy axis because its total energy is the lowest in the case of the H VSe$_2$, but the y (armchair) axis is the easy axis for the H' VSe$_2$.

\begin{table}[h]
    \caption{\label{table4}%
        The total energies of the H and H' structures of VSe$_2$ under different strains, with the magnetic moment orienting in the [100], [010], and [001] directions.   }
    \begin{ruledtabular}
        \begin{tabular}{cccc}
            \textrm{Phase} & \textrm{Strain} & \textrm{Direction} & \textrm{Total energy (eV)} \\ \colrule
            H     & $b_0=11.64$ \AA{} ($\eta_{y}$=0)   & [100] & -64.9957 \\
                  &  & [010] & -64.9956 \\
                  &  & [001] & -64.9930 \\
                  & $\eta_{yc1}$=10.1\%    & [100] & -64.2093 \\
                  &  & [010] & -64.2087 \\
                  &  & [001] & -64.2079 \\
                  & $\eta_{yc2}$=12.7\%    & [100] & -63.9099 \\
                  &  & [010] & -63.9089 \\
                  &  & [001] & -63.9083 \\ \hline
            H'    & $b_0^\prime$=12.14 \AA{} ($\eta_{y}$=4.3\%) & [100] & -64.3656 \\
                  &   & [010] & -64.3660 \\
                  &   & [001] & -64.3636 \\
                  & $\eta^\prime_{yc1}$=5.6\% ($\eta_{yc1}$)& [100] & -64.1054 \\
                  &  & [010] & -64.1057 \\
                  &  & [001] & -64.1037 \\
                  & $\eta^\prime_{yc2}$=8.1\% ($\eta_{yc2}$) & [100] & -63.9043 \\
                  &  & [010] & -63.9045 \\
                  &  & [001] & -63.9026 \\
        \end{tabular}
    \end{ruledtabular}
\end{table}

\begin{figure*}
    \centering
    \includegraphics[width=\textwidth]{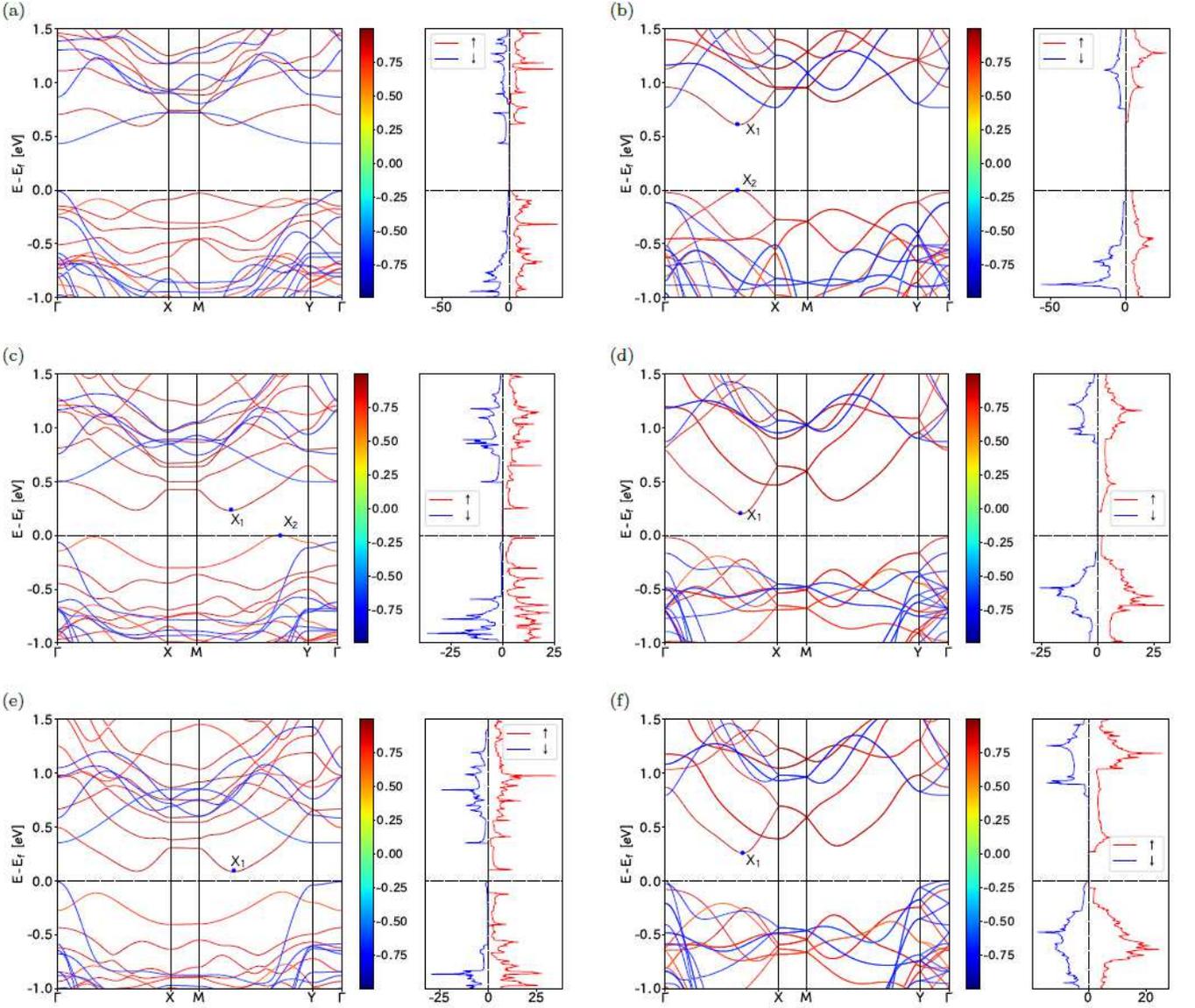}
    \caption{\label{fig5} The spin-resolved band structures and densities of states (DOSs) of VSe$_2$ in the H' structure without strain (a), with $\eta^\prime_{yc1}=5.6$\% (c), and with $\eta^\prime_{yc2}=8.1$\% (e); and in the H structure without strain (b), with $\eta_{yc1}=10.1$\% (d), with $\eta_{yc2}=12.7$\% (f). The color bar describes the electronic spin polarization, where $\pm1$ corresponds to the two 100\% spin polarizations.}
\end{figure*}

These indicate that for the H and H' structures the direction of the magnetic moment is in the xy plane and there are three magnetic easy-axe. This can be compared to three magnetic easy-axe in three-dimensional cubic ferromagnetic systems. Thus, the magnetic behavior can be described by a Heisenberg spin model with single-ion anisotropy. These mean that there can be a no-zero transition temperature $T_c$ because the Mermin-Wagner theorem will not hold due to the easy-axis anisotropy. The Hamiltonian of this model reads
\begin{equation}
    H = -\frac{1}{2}\sum_{ij}J_{ij}\bm{S}_{i}\cdot \bm{S}_{j} - A \sum_i (S^{z}_{i})^2,
    \label{eq:eq1}
\end{equation}
where $J$ is the nearest neighbor exchange interaction and $A$ describes the magnetic anisotropy. For such spin models, it is convenient to use a reliable empirical formula based on Monte Carlo simulations in order to evaluate the Curie temperature\cite{r30},
\begin{equation}
    T_c=T^{\rm Ising}_{c}\tanh ^{\frac{1}{4}}\left[\frac{6}{N_{nn}}\ln\left(1+\gamma \frac{J}{A}\right)\right],
    \label{eq:eq2}
\end{equation}
where $N_{nn}$ is the number of the nearest neighbors and $\gamma = 0.033$, and $T^{\rm Ising}_{c}$ is the critical temperature for the corresponding Ising model. $T^{\rm Ising}_{c}$ can be written as $T^{\rm Ising}_{c} = S^{2}J\tilde{T}_c /k_{B}$, where $\tilde{T}_c$ is a dimensionless critical temperature, equaling 3.64 for hexagonal lattice. For the H structure of VSe$_2$, the  nearest neighbor exchange interaction $J$ is 38.63 meV, which is calculated from the energy difference between the AFM and FM phase. The magnetic anisotropy $A$ is 0.11 meV, and thus the $T_c$ is equivalent to 87.4 K. For the H' structure and the H structure with the uniaxial strain, although the hexagonal symmetry is broken, an estimated effective $J$ in terms of this model can be useful to understand the magnetic behavior of the H structure with uniaxial strain and H' structure. Our calculation indicates that the MAE of the H structure for $\eta_{yc2}=12.7$\% (the transition y strain for the uniaxial stress) is 0.96 meV. If the effective $J$ is not changed substantially by $\eta_y$, the $T_c$ will become 154.0 K, which is higher than that of the strain-free H structure. If the effective $J$ of the H' VSe$_2$ for the same $\eta_{yc2}=12.7$\% is reduced to half of the $J$ of the strain-free H structure, the $T_c$ will still be 94.0 K. Therefore, the uniaxial strain or stress can enhance the $T_C$ values for H-phase VSe$_2$, and the $T_C$ of the H' phase should be much higher than that of the strain-free H phase.

\begin{table*}[ht]
    \begin{ruledtabular}
        \caption{\label{table5} Anisotropic effective electron masses ($m^e$) and effective hole masses ($m^h$) of VSe$_2$ in the H and H' phases at the equilibrium lattice constants and the two transition points ($\eta_{yc1}$ and $\eta_{yc2}$). $m_0$ is the free electorn mass.}
        \begin{tabular}{cccccccc}
        Phase & strain  & VBM  & CBM  & $m_x^{h}$ (VBM) & $m_y^{h}$ (VBM) & $m_x^{e}$ (CBM) & $m_y^e$ (CBM) \\          \hline
            H  & $b_0=11.64$\AA{}  ($\eta_{y}$=0)& $X_2$    & $X_1$    & 0.088$m_0$   & 0.087$m_0$   & 0.075$m_0$  & 0.075$m_0$  \\
               & $\eta_{yc1}$=10.1\%       & $\Gamma$ & $X_1$    & 0.150$m_0$   & 0.299$m_0$   & 0.121$m_0$    & 0.053$m_0$   \\
               & $\eta_{yc2}$=12.7\%       & $\Gamma$ & $X_1$    & 0.305$m_0$   & 0.013$m_0$   & 0.269$m_0$    & 0.054$m_0$   \\ \hline
            H' & $b^\prime_0=12.14$\AA{} ($\eta^\prime_{y}$=0, $\eta_{y}$=4.3\%) & $\Gamma$ & $\Gamma$ & 0.113$m_0$ & 0.051$m_0$  & 1.827$m_0$  & 0.217$m_0$ \\
               & $\eta^\prime_{yc1}$=5.6\% ($\eta_{yc1}$) & $X_2$    & $X_1$    & $\infty$ & 0.114$m_0$   & $\infty$  & 0.113$m_0$ \\
               & $\eta^\prime_{yc2}$=8.1\% ($\eta_{yc2}$) & $\Gamma$ & $X_1$    & 0.271$m_0$    & 0.034$m_0$   & $\infty$  & 0.103$m_0$  \\
        \end{tabular}
    \end{ruledtabular}
\end{table*}

\subsection{Electronic structures and band-edge properties}

The main electronic structures of VSe$_2$ under the uniaxial strain or stress are shown in Fig. \ref{fig5}. The H' phase is a ferromagnetic direct-gap semiconductor, and both CBM and VBM are at the $\Gamma$ point. When uniaxial strain or stress is applied, the CBM and VBM will shift in the Brillouin zone, with the CBM at $X_1$ between Y and M points, and the VBM at $X_2$ between Y and M points for $\eta^\prime_{yc1}$=5.6\% and at $\Gamma$ for $\eta^\prime_{yc2}$=8.1\%. It is clear that the semiconductor gap of the H' structure decreases with uniaxial strain or stress. The gap is 0.441 eV for the strain-free case, 0.249 eV for $\eta^\prime_{yc1}$=5.6\%, and 0.098 eV for $\eta^\prime_{yc2}$=8.1\%. The H phase is also a ferromagnetic direct-gap semiconductor, with the CBM and VBM at $X_1=X_2$=(0.3333,0) in the k space. The uniaxial strain or stress will make the VBM transit to $\Gamma$ point, leaving the CBM at $X_1$. The gap of H structure is 0.608 eV for the strain-free case, 0.218 eV for $\eta_{yc1}$=10.1\%, and 0.255 eV for $\eta_{yc2}$=12.7\%.

\begin{figure*}
    \includegraphics[width=0.9\textwidth]{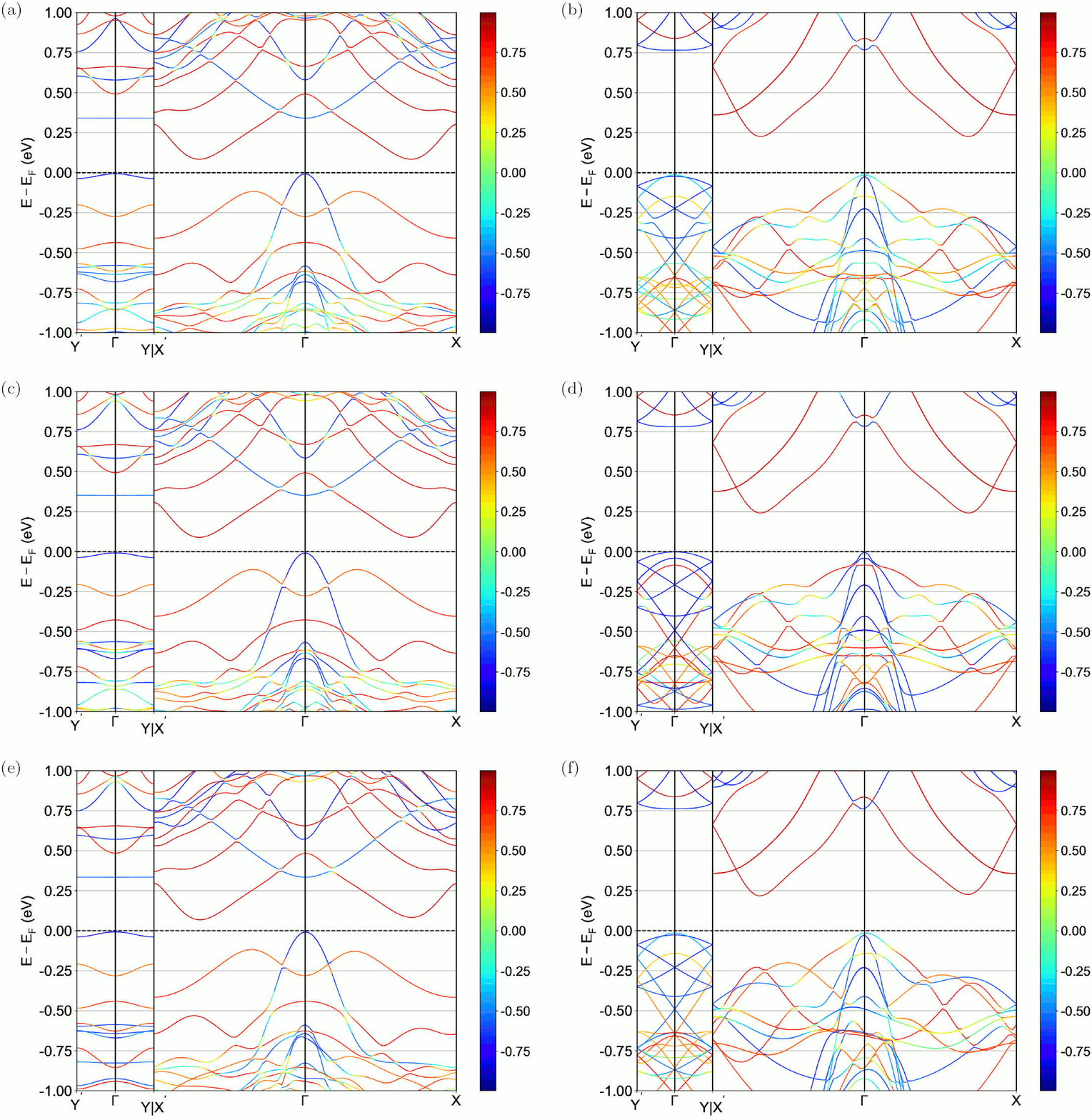}
    \caption{\label{fig6} The spin-resolved band structures of the H'-phase VSe$_2$ with $\eta_{yc2}=12.7$\% with the magnetic moment orienting in the x axis (a), the y axis (c), and the z axis (e); and those of the H-phase VSe$_2$ in the three axes (b, d, f). The color bar describes the corresponding electron spin polarization. }
\end{figure*}

The uniaxial strain and stress can also change the effective masses of the VBM and CBM in the k-space, as shown in Table~\ref{table5}. For the H structure without strain, the effective masses of the VBM and CBM are isotropic. When the uniaxial strain or stress is applied, the effective carrier masses of the VBM and CBM become anisotropic because of the broken symmetry. The effective electron mass along the x-axis, $m_x^{e}$, becomes larger by 72\% (or 248\%) than that of the strain-free case, and $m_y^e$ along the y-axis is multiplied by 3.42 (or 0.15 ) for $\eta_{yc1}$=10.1\%  (or $\eta_{yc2}$=12.7\%). Under the uniaxial stress or strain, the effective hole mass is also changed. For $\eta_{yc1}$, the x-axis and y-axis effective hole masses ($m_x^{h}$ and $m_y^{h}$) become larger than those without any strain, and the y-axis effective hole mass $m_y^{h}$ is larger than x-axis effective hole mass $m_x^{h}$. The y-axis effective hole mass of VBM under $\eta_{yc2}$=12.7\% (uniaxial stress) is much smaller than the effective hole mass without any stress.

For the H' structure, the situation is also complicated. The VBM is located at $\Gamma$ for zero strain and $\eta^\prime_{yc2}$, but it is at the X$_2$(0.127,0.5) between the Y and M points in the Brillouin zone for $\eta^\prime_{yc1}$. Actually, the point X$_3$(0.127,0) between the $\Gamma$ and X points is almost degenerate to X$_2$, with a small energy difference 0.0371eV, and thus the effective hole mass $m^h_x$ along the X$_2$-X$_4$ line (the x-direction)  is very large (infinity), which is consistent with the peak in the DOS. As for the CBM, the uniaxial strain or stress makes it transit from $\Gamma$ to the X$_1$(0.34745,0.5) which is also located at the boundary of the Brillouin zone. The energy band near  X$_1$ (in the x direction) is also flat, with the maximal energy difference 0.0048eV for $\eta^\prime_{yc1}$=5.6\% and 0.0043eV for $\eta^\prime_{yc2}$=8.1\%, and therefore the effective electron masses $m_x^{e}$ at the CBM are nearly infinity for $\eta^\prime_{yc1}$ and $\eta^\prime_{yc2}$.

\subsection{Comparison of the two phases}

The electronic structure  depends on the direction of the magnetic moment because of the spin-orbit coupling. The main properties of band structures of the H' and H VSe$_2$ under $\eta_{yc2}$=12.7\%  (the transition point for the uniaxial stress), with the magnetic moment orienting in [100], [010], and [001] directions, are shown in Fig.~\ref{fig6}. The effective hole masses of the VBM are shown in Table~\ref{table6}. For the H structure with $\eta_{yc2}$=12.7\%, the easy-axis is in the x-axis because the energy difference between the [010] and [100] axe is 0.96 meV, but for the H' structure with the same $\eta_{yc2}$, the wasy-axis is in the y-axis. For both of the structures, the y-axis effective hole mass $m^h_y$ is much larger than the x-axis effective hole mass $m^h_x$. Furthermore, it is clear that the spin polarization of the VBM is also affected by the direction of the magnetic moment. Very interestingly, for the H phase, the effective hole mass $m^h_x$ ($m^h_y$) can be switched from 0.076$m_0$ to 0.014$m_0$ (from 0.132$m_0$ to 0.313$m_0$) by switching the magnetic moment from the easy-axis to the y-axis; and for the H' structure, $m^h_x$ remains isotropic and $m^h_y$ can be enhanced by 7\% through switching the magnetic moment from the easy-axis to the x-axis. It is indicated that the properties of carriers near the VBM can be controlled by switching the direction of the magnetic moment through applying external magnetic field.

\begin{table}[h]
    \caption{\label{table6}The effective hole masses of the VBM of the H and H' phases of VSe$_2$ with $\eta^\prime_{yc2}$=8.1\% ($\eta_{yc2}=12.7$\%) with the magnetic moment orienting in the [100], [010], and [001] directions. $m_0$ is the free electorn mass.}
    \begin{ruledtabular}
        \begin{tabular}{cccc}
            Phase & direction &  $m^h_x$ (x axis) & $m^h_y$ (y axis)\\ \hline
            H   & [100]   & 0.076$m_0$       & 0.132$m_0$       \\
                & [010]   & 0.014$m_0$       & 0.313$m_0$       \\
                & [001]   & 0.076$m_0$       & 0.132$m_0$       \\ \hline
            H'  & [100]   & 0.034$m_0$       & 0.287$m_0$       \\
                & [010]   & 0.034$m_0$       & 0.307$m_0$       \\
                & [001]   & 0.034$m_0$       & 0.287$m_0$       \\
        \end{tabular}
    \end{ruledtabular}
\end{table}

\begin{table}[h]
    \caption{\label{table7}%
        The bond lengthes $l_b$ (\AA) and angles $\alpha_b$ ($^\circ$) of the H and H' phases of VSe$_2$ at $\eta_{yc2}=12.7$\% ($\eta^\prime_{yc2}$=8.1\%). The V-Se and V-Se' share the same bond length.}
    \begin{ruledtabular}
        \begin{tabular}{ccc|ccc}
        $l_b$ & \textrm{H' phase} & \textrm{H phase} & $\alpha_b$ & \textrm{H' phase} & \textrm{H phase}                                      \\ \colrule
            V1-Se1  & 2.53 & 2.53 & Se1-V1-Se1' & 77.09    & 77.17\\
            V1-Se2  & 2.63 & 2.68& Se2-V1-Se2' & 75.64    & 72.03\\
            V2-Se2  & 2.51 & 2.53 & Se2-V2-Se2' & 79.69    & 77.17\\
            V2-Se3  & 2.76 & 2.68& Se3-V2-Se3' & 55.22    & 72.03\\
            V3-Se3  & 2.59 & 2.53 & Se3-V3-Se3' & 59.25    & 77.17 \\
            V3-Se4  & 2.54 & 2.68& Se4-V3-Se4' & 79.29    & 72.03\\
            V4-Se4  & 2.55 & 2.53 & Se4-V4-Se4' & 79.05    & 77.17\\
            V4-Se1  & 2.60 & 2.68& Se1-V4-Se1' & 74.47    & 72.03\\ \hline
            Se1-Se1' &3.15&3.15  &  & & \\
            Se2-Se2' &3.23&3.15  &  & & \\
            Se3-Se3' &2.56&3.15  &  & & \\
            Se4-Se4' &3.24&3.15  &  & &\\
        \end{tabular}
    \end{ruledtabular}
\end{table}

\begin{figure}
    \includegraphics[width=0.48\textwidth]{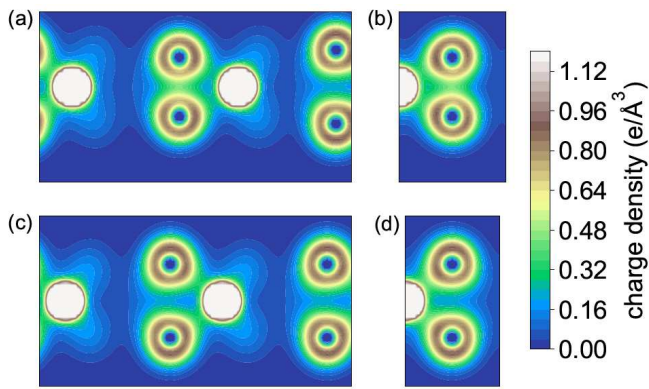}
    \caption{\label{fig7} The 2D charge density of the H'-phase VSe$_2$ with $\eta_{yc2}$\%  in the plane of Se3-V2-Se3' (a) and Se3-V3-Se3' (b); and those of the H-phase VSe$_2$ with $\eta_{yc2}$ in the  two similar planes (c, d). The color bar describes the charge density, with the increment 0.04 $e/\text{\AA}^3$.}
\end{figure}

The bond lengthes and band angles of the H' and H phases of VSe$_2$ under $\eta_{yc2}=12.7$\% ($\eta^\prime_{yc2}$=8.1\%) are summarized in Table \ref{table7}. It is clear that the V-Se and Se-Se bond lengthes are different between the H and H' structures. The largest difference appears in the V3-Se4 bond length which becomes smaller by 5\% if switching from the H to H' structure. As for the Se-Se bond lengthes, when switching from H to H' structure, the vertical Se3-Se3' bond is shortened by 19\% and the Se4-Se4' bond length is enhanced by 3\%. When the H' structure is achieved from the H structure, the bond angles Se3-V2-Se3' and Se3-V3-Se3' become smaller by 23\% and the bond angle Se4-V3-Se4' is enhanced by 10\%. It becomes clear that the drive force for forming the H' phase can be attributed to the substantial reduction of vertical Se3-Se3' bond length. On the other hand, we present in Fig.~\ref{fig7} the 2D charge density of the two phases in the planes defined by Se3-V2-Se3' and Se3-V3-Se3'. These charge density shows that there are strong covalent bond between Se3 and Se3' in the H's structure, in contrast to that in the H phase. Remembering also that the V atoms near the Se3-Se3' pair contribute one more $\mu_B$ to the magnetic moment in the H' phase, we can conclude that the covalent Se3-Se3' bond makes the nearby V atoms assume 3+ instead of 4+. All these factors lead us to the conclusion that the structural and bond reconstruction is formed in the H' phase of VSe$_2$.

\section{Conclusion}

Through first-principles investigation, we have found a metastable ferromagnetic semiconductive H' phase in H VSe2 and some other similar, and shown that the new phase can appear when these H monolayer materials are under uniaxial strain or uniaxial stress along the armchair (y) direction. For the uniaxial stress (uniaxial strain) scheme, the H' phase will become lower in total energy than the H phase at the transition strain point $\eta_{yc2}$ ($\eta_{yc1}$). The calculated phonon spectra indicate the dynamical stability of the H' structure of VSe$_2$, VS$_2$, and CrS$_2$, and the path of phase switching between the H and H' structures of VSe$_2$ is calculated, obtaining the energy barrier 0.012 eV. For VSe$_2$, the easy magnetic axis for the H' phase is in the y direction, in contrast to the x direction for the H phase, and the H' phase has stronger ferromagnetism and much higher Currier temperature than the H phase. The Currier temperature can be substantially enhanced by applying uniaxial strain or stress, especially transiting to the H' phase.

Spin-resolved electronic structures, energy band edges, and effective carrier masses for both of the ferromagnetic semiconductive H and H' phases can be substantially changed by the applied uniaxial strain or stress, leading to some flat bands (huge effective masses) near the band edge of the strained H' phase. Our calculation with the spin-orbit coupling shows that the valence band edges are dependent on the magnetic moment orientation, which implies that the effective hole masses can be manipulated by switching the magnetic moment orientation away from the easy axis. For the H phase with $\eta_{yc2}=12.7$\% for the uniaxial stress scheme, when the magnetic moment is switched to the y axis, the anisotropic effective hole mass ($m^h_x$,$m^h_y$)=(0.076$m_0$,0.132$m_0$) can be changed to (0.014$m_0$,0.313$m_0$), with the anisotropy ratio $m^h_y/m^h_x$ jumping from 1.7 to 22.4.

As for the bond length difference between the H' and H phases, the largest one is -17\% or -19\% ($b=b^\prime_0$ or $b_{c2}$) for the Se3-Se3' bond. The largest bond angle difference reaches to -20\% or -23\% ($b=b^\prime_0$ or $b_{c2}$) in the cases of the Se3-V2-Se3' and Se3-V3-Se3' bond angles. The charge density in the planes of Se3-V2-Se3' and Se3-V3-Se3' clearly shows noticeable covalent bond between Se3 and Se3' in the H' phase. Therefore, the substantially shortened Se3-Se3' bond, corresponding to noticeable covalent bond, switches the valence of the nearby V atoms, leading to the enhanced ferromagnetism and structural reconstruction. Hence, the structural and bond reconstruction can be realized by applying uniaxial stress or strain  in 2D H VSe$_2$ and some other similar. These can stimulate more exploration in such 2D ferromagnetic semiconductive materials to seek more insights, interesting phenomena, and potential applications.

\begin{acknowledgments}
This work is supported by the Nature Science Foundation of China (Grant Nos.11974393 and 11574366) and the Strategic Priority Research Program of the Chinese Academy of Sciences (Grant No. XDB33020100). All the numerical calculations were performed in the Milky Way \#2 Supercomputer system at the National Supercomputer Center of Guangzhou, Guangzhou, China.
\end{acknowledgments}

\bibliographystyle{apsrev4-1}


\end{document}